\documentclass[fleqn,10pt]{wlscirepMain}
\usepackage{subfigure}

\newcommand{\g}[1]{\ensuremath{\mathbf{#1}}}
\newcommand{\ket}[1]{\ensuremath{\vert #1 \rangle}}

\newcommand{\psii}{\ensuremath{\psi_{0i}(\g{r})}}

\newcommand{\thi}{\ensuremath{\theta_{i}(\g{r})}}
\newcommand{\psijc}{\ensuremath{\psi_{0j}^{*}(\g{r})}}
\newcommand{\psij}{\ensuremath{\psi_{0j}(\g{r})}}

\newcommand{\zic}{\ensuremath{\hat \zeta_{i}^{\dag}(\g{r})}}
\newcommand{\zi}{\ensuremath{\hat \zeta_{i}(\g{r})}}
\newcommand{\zjc}{\ensuremath{\hat \zeta_{j}^{\dag}(\g{r})}}
\newcommand{\zj}{\ensuremath{\hat \zeta_{j}(\g{r})}}

\title{Tunable Polarons in Bose-Einstein Condensates}

\author[1]{E. Compagno}
\author[2,*]{G. De Chiara}
\author[3,4]{D.G. Angelakis}
\author[5]{G. M. Palma}
\affil[1]{Department of Physics and Astronomy, University College London, Gower Street, WC1E 6BT London, United Kingdom.}
\affil[2]{Centre for Theoretical Atomic, Molecular and Optical Physics Queen's University, Belfast BT7 1NN, United Kingdom.}
\affil[3]{School of Electronic and Computer Engineering, Technical University of Crete, Chania, Crete, 73100 Greece.}
\affil[4]{Centre for Quantum Technologies, National University of Singapore, 2 Science Drive 3, 117542 Singapore.}
\affil[5]{NEST-INFM (CNR) and Dipartimento di Fisica e Chimica Universit\`a degli Studi di Palermo, Via Archirafi 36, I-90123 Palermo, Italy.}
\affil[*]{Correspondence and requests for materials should be addressed to G. De Chiara (email: g.dechiara@qub.ac.uk)}

\begin{abstract}
A toolbox for the quantum simulation of polarons in ultracold atoms is presented. Motivated by the impressive experimental advances in the area of ultracold atomic mixtures, we theoretically study the problem of ultracold atomic impurities immersed in a Bose-Einstein condensate mixture (BEC). The coupling between impurity and BEC  gives rise to the formation of polarons whose mutual interaction can be effectively tuned using an external laser driving a quasi-resonant Raman transition between the BEC components. Our scheme allows one to change the effective interactions between polarons in different sites from attractive to zero. This is achieved by simply changing the intensity and the frequency of the two lasers. Such arrangement opens new avenues for the study of strongly correlated condensed matter models in ultracold gases.  
\end{abstract}
\begin{document}

\flushbottom
\maketitle

\thispagestyle{empty}

\noindent 
\section*{Introduction}
Ultracold atomic gases are one of the most advanced experimental and theoretical platform for quantum simulations \cite{trabesinger_quantum_2012-1}, ranging from continuous systems in shallow traps to discrete models on a lattice \cite{lewenstein_ultracold_2012,bloch_quantum_2012}, with either short contact or long range dipolar interactions, with various  dimensionalities and in the presence of impurities or disorder \cite{dutta_non-standard_2015}.
Such flexibility makes possible the simulation of typical models of condensed matter physics, high energy physics, quantum biology and chemistry ~\cite{hild_far--equilibrium_2014,fukuhara_microscopic_2013,fukuhara_quantum_2013,buchler_atomic_2005,jaksch_creation_2002,ferlaino_efimov_2011}. 
Perhaps even more importantly,  they allow the simulation of quantum systems in ranges of parameters which go beyond standard models of condensed matter physics. A paradigmatic examples is provided by polarons, introduced in condensed matter to describe charge carriers dressed by lattice phonons , the standard example being  electrons moving in metals. Although bosonic carriers are unusual in solid state systems, ultracold bosonic atoms in optical lattices are the perfect platform for studying bosonic polarons. 
Ultracold polarons are indeed a well established area of  both theoretical and experimental investigation
Triggered by experimental advances in cold atomic mixtures and impurities immersed in Bose-Einstein condensates (BEC) \cite{spethmann_dynamics_2012,catani_quantum_2012,scelle_motional_2013,hu_bose_2016-1}, many theoretical works have analysed interesting problems in this context. These include the study of effective polaron-polaron interactions \cite{klein_interaction_2005,cucchietti_strong-coupling_2006,bruderer_polaron_2007,posazhennikova_two_2009-1,rath_field-theoretical_2013}, clustering and transport of polarons \cite{klein_dynamics_2007
,bruderer_transport_2008,johnson_impurity_2011,grusdt_bloch_2014,grusdt_renormalization_2015}, self-trapping of impurities\cite{bruderer_self-trapping_2008}, multi-polaron problems \cite{casteels_many-polaron_2011,santamore_multi-impurity_2011,casteels_bipolarons_2013}, probing BEC with impurities \cite{bruderer_probing_2006,ng_single-atom-aided_2008,cirone_collective_2009,sabin_impurities_2014}, BEC-generated entanglement of impurities \cite{mcendoo_entanglement_2013}, non-Markovian environments \cite{haikka_quantifying_2011}. Furthermore attractive and repulsive polarons have been theoretically studied \cite{li_variational_2014,massignan_polarons_2014,ardila_impurity_2015} and, in the fermionic case, experimentally observed \cite{kohstall_metastability_2012,koschorreck_attractive_2012}. A two-band model for the impurity has been recently considered in \cite{yin_polaronic_2015}. 

 An important feature of BEC polarons is that when more than one impurity is immersed in a BEC  an effective boson-mediated interaction between polarons appears \cite{bruderer_transport_2008,naidon_two_2016}. Indeed the presence of an impurity atom deforms locally the BEC equilibrium wavefunction, inducing an evironment-mediated interaction between polarons, whose strength depends on the physical characteristics both of the impurity atoms and of environment itself. 

Up to now, the environment considered for the BEC-polaron problem consists of a single component BEC. However so far the physics of polarons in composite reservoirs, i. e. consisting of BEC mixtures, has not been investigated. Although composite systems of interacting BEC have been studied \cite{hong_collision_2001,leblanc_species-specific_2007,klaiman_solvable_2016,barfknecht_dynamical_2016}, the effect of their coupling with an impurity has not yet been addressed.

Scope of the present manuscript is to show how substantially new features can be engineered in experimentally accessible ultracold atomic polarons if one considers a BEC mixture of two Raman coupled hyperfine states. Our system consists of impurity atoms immersed in a two component BEC. We will show how  the polaron  parameters and the inter-polaron interaction can be modified  by tuning the Raman coupling between the BEC hyperfine levels of the BEC. 
In particular we show that the inter-polaron interaction can, in our model, be tuned without changing the impurity BEC coupling strength but just the internal Raman coupling.  We first analytically calculate the deformation of the two-component BEC induced by the presence of the impurities. We then introduce the Raman coupling and calculate the effective interaction potential between two polarons and show how it can be tuned by varying the Raman coupling strength.
Our scheme appear to be within the actual experiment feasibility using single atom addressing techniques \cite{weitenberg_single-spin_2011,preiss_strongly_2015} and a species-selective dipole potential \cite{catani_quantum_2012,leblanc_species-specific_2007} in a two photons Raman coupling scheme \cite{galitski_spin-orbit_2013}. 
In the following section we will illustrate our main results, all the details about the model Hamiltonian and the derivation being described in Methods.

\section*{Results}
\begin{figure}[t]
\centering
\includegraphics[width= 0.6\columnwidth]{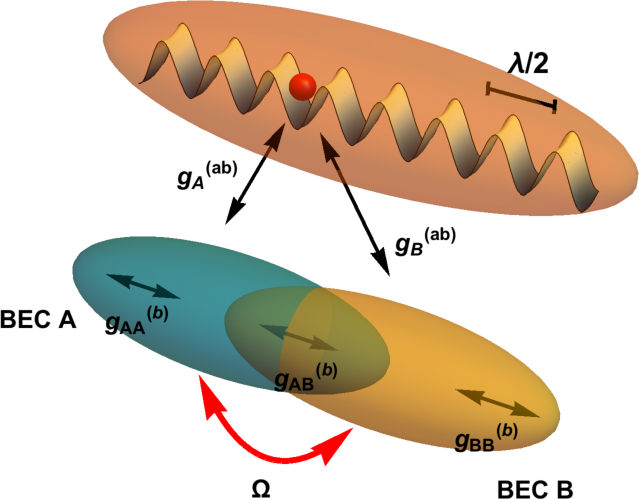} 
\caption{ \textbf{Scheme of the model}: an impurity atom (red), trapped in a fixed site of a deep optical lattice with spacing $a=\lambda{/}2$ (orange), interacts with a mixture of two BECs (green and yellow) trapped in a very shallow potential. The collisional coupling strengths between pairs of atoms of the two components of the condensates and  between impurities and BEC atoms are respectively $g_{ij}^{(b)}$ and $g_{i}^{(ab)}$. The two BECs are also coupled via a two-photon Raman term $\Omega$.}
\label{fig:2BECModelScheme}
\end{figure}
\subsection*{Effective deformations induced by impurity atoms}
The system, sketched in Fig. \ref{fig:2BECModelScheme}, consists of impurity atoms trapped in an optical lattice, (orange) interacting with a two component BEC. Specifically the mixture is made of a single atomic species in two different hyperfine states (the two components are represented respectively in yellow and green in Fig. \ref{fig:2BECModelScheme}). Such configuration is experimentally realisable by using species-selective optical potentials, namely an optical lattice experienced only by the impurity atoms in the system \cite{catani_quantum_2012,leblanc_species-specific_2007}. The average heating of this kind of configuration depends strongly on the detuning of the trapping laser from the resonant atomic transition, and it has been quantitatively evaluated in Ref. \cite{leblanc_species-specific_2007}.

As shown in Methods, we assume that the two component BEC is trapped in a potential shallow enough 
to assume a uniform free condensate density. Indeed the curvature of the trapping potential in typical BEC experiments \cite{gerbier_interference_2005} can be, with a good approximation, be neglected in the centre of the trap. Besides, spatially uniform condensates have been experimentally realised in an optical box-like trap \cite{gaunt_bose-einstein_2013}. 

All the atoms in the system weakly interact with each other via a repulsive contact pseudo-potential. 
Specifically, in Fig. \ref{fig:2BECModelScheme}, we indicate with $g_{ij}^{(b)}$ the collisional coupling constants between a pair of atoms of the two component BEC and with $g_{j}^{(ab)}$ the coupling constants between an impurity and one atom of the $j$ component of the BEC mixture ($j\in \left\{A,B\right\}$ as indicated in Fig. \ref{fig:2BECModelScheme}). As discussed in Methods we also include an external laser coupling between the two BEC hyperfine levels $\ket{a}$ and $\ket{b}$, $\Omega$ in Fig. \ref{fig:RamanTransition}, leading to an effective Raman coupling $\hat H_{\text{Ram}}$,
as shown in \cite{kanamoto_quantum_2007,brion_adiabatic_2007,cirac_quantum_1998,jaksch_cold_2005}. 

Let us start by considering static impurities trapped, in the tight-binding limit, in a deep optical lattice potential with negligible hopping. The presence of an impurity atom creates a localised spatial deformation of the condensate order parameters giving rise to a static polaron. 

We characterise the structure of such polaron via a mean field approach, (see Methods for details) namely substituting the BEC field operators $\hat \psi_i(\g{r})$ with the corresponding order parameters $\psi_i(\g{r})$. These in turn can be expressed as the sum of an unperturbed uniform $\psi_{0i}(\g{r})$ term plus a linear correction term $\theta_i(\g{r})$ which describes the BEC deformation induced by the impurities \cite{bruderer_polaron_2007,bruderer_transport_2008,cucchietti_strong-coupling_2006}. This allows us to obtain the Gross-Pitaevskii equations for the order parameters in presence of impurities.

Using a standard approach we first neglect the presence of the impurity atoms $g_i^{(ab)}{=}0$ (and therefore we assume 
$\theta_i(\g{r}){=}0$). For a uniform condensate $\psi_{0i}(\g{r}){=}\sqrt{n_{0i}}\ e^{i \g{q}_i \cdot \g{r}}$ \cite{fetter_nonuniform_1972}, and the stationary condition $\delta H_{GP}/ \delta \psi_{0i}^*{=}0$ leads to the following expression for the chemical potential  $\mu_i^{(b)}{=}2g_{ii}^{(b)} n_{0i}{+}g_{i\neq j}^{(b)} n_{0j} {-}\frac{\hbar\Omega}{2}\sqrt{\frac{n_{0j}}{n_{0i}}}$, where $\Omega$ is the Raman coupling strength defined in equation \eqref{eq:LaserRamanHam} (see Methods for details). The next step is to take into account the presence of the impurity and  to impose stationary conditions $\delta H_{GP}/\delta \theta_i{=}0$. Assuming $\theta_i(\g{r})$ to be real, these lead to the following coupled differential equations 
\begin{equation} 
\left[\nabla^2 -\mathbb{M}\right]
\begin{pmatrix}
\theta_A(\g{r})\\
\theta_B(\g{r})\\
\end{pmatrix}
=
\begin{pmatrix}
\gamma_A\rho(\g{r})\\
\gamma_B\rho(\g{r})\\
\end{pmatrix},\ 
\label{eq:MatrixFormSysDef2BECModel}
\end{equation}
where $\mathbb{M}{\equiv} \left(\mathcal{M}_{ij}\right)$, ($i,j{=}A,B$) and $\rho(\mathbf{r})$ is the average density of impurities (see Methods for details). We find 
\begin{equation}
\begin{split}
\mathcal{M}_{ii}&=\frac{2 m_b}{\hbar^2}\left[4 g_{ii}^{(b)} n_{0i}+\frac{\hbar\Omega}{2}\sqrt{\frac{n_{0j}}{n_{0i}}}\right]~,\\
\mathcal{M}_{i\neq j}&=\frac{2 m_b}{\hbar^2}\left[2 g_{i\neq j}^{(b)} \sqrt{n_{0i}}\sqrt{n_{0j}}-\frac{\hbar\Omega}{2}\right]~,\\
\gamma_i&=\frac{2 m_b}{\hbar^2}\left[g_{i}^{(ab)}\sqrt{n_{0i}}\right].
\end{split}
\label{eq:CoeffabcLaser}
\end{equation}
Note that the off diagonal terms $\mathcal{M}_{i\neq j}$ are due to the coupling  between the hyperfine sub levels $g_{i\neq j}^{(b)} $. We mention that the previous system can be easily generalised to a non-uniform trapping potential. In this case it can be solved via numerical methods.

\subsection*{Polarons in two components BEC in the absence of Raman Coupling}
In the absence of Raman coupling ($\Omega{=}0$), the system of differential equations
(\ref{eq:MatrixFormSysDef2BECModel}) can be decoupled by expressing the deformations
$\theta_i$ in terms of new effective deformations $\theta_i'$, defined in equation  \eqref{eq:EffDeformTransform} in Methods. These are described by two separate modified Helmholtz equations whose solutions, expressed in terms of Green functions are
\begin{equation}
\theta_{\pm}'(\g{r})=\mathcal{K}_{\pm} \int d\g{r}'\ \mathcal{G}(\g{r}-\g{r}',\eta_{\pm})\rho(\g{r}'), 
\label{eq:SolEffectiveDefModel}
\end{equation}
where the explicit form of the constants $\mathcal{K}_{\pm}$ is given in equation  \eqref{eq:KConstantsNoLaser}. Hereafter we assume both the optical lattice and the two component BEC to be one-dimensional. 
In this case the Green function is  $\mathcal{G}(x-x',\eta_{\pm}){\equiv} \frac{1}{2\eta_{\pm}}\exp\left(-\eta_{\pm}\vert x{-}x'\vert\right)$ \cite{arfken_mathematical_2005} and, assuming equal densities for the hyperfine BEC levels   (i.e. $n_{0A}{=}n_{0B}{=}n/2$) the effective healing lengths take the form
\begin{equation}
\eta_{\pm}=\left[ \left(\frac{2m_b n}{\hbar^2}\right)
\left(g_{AA}^{(b)}+g_{BB}^{(b)}
{\pm}\sqrt{(g_{AA}^{(b)}-g_{BB}^{(b)})^2 + (g_{AB}^{(b)})^2}\right)\right]^{1/2}
\label{eq:MEtaAEtaBnAnB}
\end{equation}
For an impurity trapped in an effective one-dimensional deep lattice $\rho (x)$ is a gaussian function and (\ref{eq:SolEffectiveDefModel}) can be analytically integrated. The explicit expressions of the effective deformations, $\theta_i'(x){=}\mathcal{K}_{i}\mathcal{F}_{\sigma,\eta_i}(x)$  ($i{=}\pm$), are given in equation \eqref{eq:FunzioniFDeforModelAppA} and equation \eqref{eq:KConstantsNoLaser} in Methods. 

To obtain quantitative results, we consider a mixture of two hyperfine states of $^{87}Rb$ with equal density $n_{0i}{=}1\ \mu m^{-1}$, interacting with impurity atoms of $^{41}K$, trapped in a deep one-dimensional optical potential. Such system has been realised in \cite{catani_quantum_2012} for a single BEC. The typical lattice spacing is $a=\lambda{/}2{=}532\ nm$ \cite{weitenberg_single-spin_2011}, which results in a recoil energy of $E_{R}{\simeq}1.34\times 10^{-30}J$. For the impurities we assume gaussian Wannier wavefunctions $\rho(x){=}\left(1/\sqrt{\pi \sigma^2}\right) \exp\left(-(x-x_0)^2/\sigma^2\right)$ whose width $\sigma$ can be tuned varying the depth of the optical potential \cite{bruderer_transport_2008}. In the following $\sigma{/}a{\simeq}0.15$, to avoid direct interaction between two impurities placed in neighbouring sites and to suppress tunnelling effects, obtained with a longitudinal trapping frequency of $\omega_{L}/2\pi{\simeq}18$kHz.
The coupling constants of the two component BEC are assumed to be $g_{AA}^{(b)}{\simeq}2.08{\times}10^{-37} J\ m$,  $g_{AB}^{(b)}{\simeq}2.03{\times} 10^{-37} J\ m$ and  $g_{BB}^{(b)}{\simeq}1.99{\times} 10^{-37} J\ m$. These values are obtained scaling the  3D scattering lengths \cite{egorov_measurement_2013,lercher_production_2011} to the 1D case following \cite{olshanii_atomic_1998} and using the transverse trapping frequency ($\omega_\perp/2\pi{\simeq}$34kHz) as in \cite{catani_quantum_2012} to ensure the one-dimensionality condition $\mu_{i}^{(b)}{/}\hbar\omega_\perp{\ll} 1$  \cite{pethick_bose-einstein_2008}. The condition $\hbar\omega_\perp {\gg} k_B T$ ensures that there are no transverse excitations due to thermal effects. In our case for $T{\simeq} 30 nK$ is $\hbar\omega_\perp/k_B T {\simeq} 10^{2}$ \cite{mazets_thermalization_2010}. We show in the Supplementary Material how the static polaron description can be further extended in presence of excitations, (\textit{e. g.} when $T{\neq}0$).

Finally the coupling constants between impurity atoms and BEC are assumed to be both of the order of $\vert g_i^{(ab)}\vert \simeq 10 g_{AA}^{(ab)}$ in agreement with values in \cite{catani_quantum_2012,thalhammer_double_2008}.

Let us stress that, as can be seen from (\ref{eq:MEtaAEtaBnAnB}), the effective deformation widths $d_i{=}1/\eta_i$ do not depend on the impurities but only on the BEC physical parameters. Indeed the presence of the impurity atom creates a localised potential acting on the - otherwise homogeneous - BEC. The size of deformation so generated is of the same order of magnitude of the healing length, as one expect, given that the Green function, whose spatial extension is the healing length, is the solution of \eqref{eq:MatrixFormSysDef2BECModel} for a delta-like source. 

On the other hand the deformation depths are proportional to $\mathcal{K}_{i}$ in equation \eqref{eq:KConstantsNoLaser}, and depend also on the coupling constants of the impurity atoms. In other words the depth of the effective deformations can be controlled for instance by using Feshbach resonances on impurity atoms (as in \cite{catani_quantum_2012}) or, limited to the static case, with confined-induced resonances \cite{massignan_three-dimensional_2006}. 

\begin{figure}[h]
\centering
\includegraphics[width= 0.3\columnwidth]{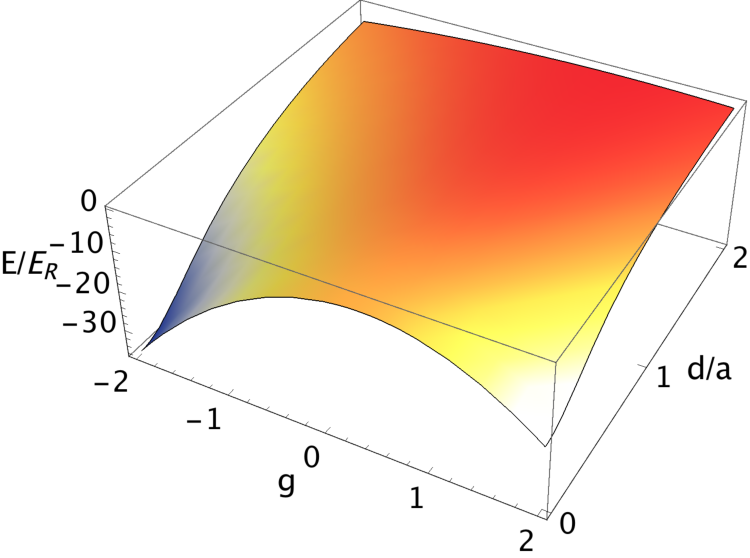} 
\includegraphics[width= 0.3\columnwidth]{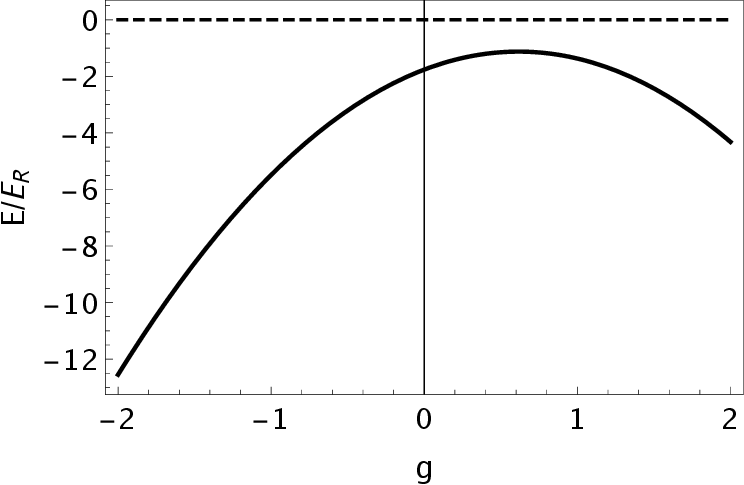} 
\includegraphics[width= 0.3\columnwidth]{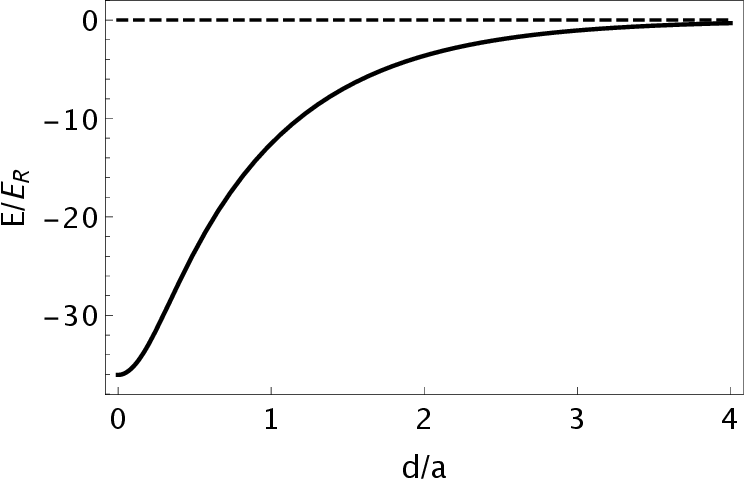} 
\caption{Energy for two impurity atoms as a function of their relative distance $d$ (in lattice spacing $a=\lambda/2$ units) and as a function of $g=g_{A}^{(ab)}/g_{B}^{(ab)}$ (left). Slice of the 3D plot at the fixed distance $d=a$ showing how the interaction between two impurities in two neighbouring sites can be modulated  by varying $g$ (centre) . 
(right) Energy for two impurity atoms as a function of their relative distance $d$ (in lattice spacing $a=\lambda/2$ units) for $g=1$.}
\label{fig:Energia2ImpurezzeMixRb87}
\end{figure}

Once we have the explicit expressions for the effective condensate deformations, we can evaluate the interaction energy between impurities and the system of BECs, using equation \eqref{eq:HGPStazCondModel}. 
The ground state energy can be written as the sum of the energy of the system of two unperturbed BECs interacting with the impurity atoms plus an additional term due to the interaction of the impurities with the order parameters' deformations. The latter allows an effective interaction between impurities mediated by the BEC. For a system of impurities immersed in the optical lattice the ground state energy is
$E_{GP}{=}A_0\sum_m n_m {+}\sum\limits_{i=\pm}\sum_{m,l}B_{i}\ G^{i}_{m,l}n_m n_l$, 
where $m,l$ run over lattice sites and  $n_m, n_l$ are the mean number of impurities and $G^i_{m,l}{\equiv}\int dx dx'\ \mathcal{G}(x-x',\eta_i)\vert \chi_{m}(x)\vert^2\vert \chi_{l}(x')\vert^2$. Here $\chi_m(x)$ are Wannier functions of the optical lattice, which shows clearly how the correction to the ground state energy is due to an effective interaction between the impurities mediated by the condensates' deformations. The expressions for coefficients $A_0$ and $B_i$ are explicitly shown in Eq.~\eqref{eq:EGPConstantiMoltipModel} of Methods. We obtain explicit analytic expressions in a deep 1D optical potential with gaussian Wannier functions. For a single impurity atom fixed at the origin of the coordinate system we find 
\begin{equation}
E_{GP}^{one}=A_0 + \sum_{i=\pm}B_i \frac{1}{2\eta_i}e^{\eta_i^2 \sigma^2 /2}\text{Erfc}\left(\frac{\eta_i \sigma}{\sqrt{2}}\right)
\label{eq:EnergiaGroundSingolaImpModel}
\end{equation}
Since $A_0$ is the ground state energy with no deformations, $E_{GP}^{one}{-}A_0{\simeq}{-}7.9 E_R$, for $g_A^{(ab)}=g_B^{(ab)}=10 g_{AA}^{(b)}$, represents the interaction energy between the impurity and the effective deformations, namely the binding energy of the polaron. In the same way one obtains the ground state energy for two impurities, fixed in $x_1$ and $x_2$ respectively, as a function of their relative distance $d{=}\vert x_1 {-} x_2\vert$. 
When $d$ is infinity the ground state energy is $2E_{GP}^{one}$, as one expects for two non interacting impurities. The variation  $ \Delta E(d)$ with respect to its value at infinite distance is an effective interaction potential between the two impurities, which turns out to be
\begin{equation}
\Delta E(d)=\sum\limits_{i=\pm}\Delta E_i(d)=\sum_{i=\pm}\frac{B_i}{4\eta_i}e^{\frac{1}{2} \eta_i \left(\sigma ^2 \eta_i -2 d\right)}\left[1{+}\text{Erf}\left(\frac{d{-}\sigma ^2 \eta_i }{\sqrt{2} \sigma }\right){+}e^{2 d \eta_i } \text{Erfc}\left(\frac{d{+}\sigma ^2 \eta_i }{\sqrt{2} \sigma }\right)\right]
\label{eq:EnergyTwoImpDistance}
\end{equation}
As it is clear in Fig. \ref{fig:Energia2ImpurezzeMixRb87} (right), where we consider two impurities with $g_j^{(ab)}=10 g_{AA}^{(b)}$, such effective interaction is attractive. We also study in Fig. \ref{fig:Energia2ImpurezzeMixRb87} (left) and (centre) how the interaction energy depends on the ratio between the coupling constants by fixing $g_B^{(ab)}=10 g_{AA}^{(b)}$ and varying $g_A^{(ab)}=g\ g_B^{(ab)}$ in the range $g\in\left[-2,2\right]$. 

Therefore we find that for two impurities immersed in a BEC mixture the effective attractive interaction can be modulated by controlling the ratio between the coupling constants between the impurity atoms and the two BEC components. We highlight that by varying the ratio between the coupling constants $g_j^{(ab)}$ one controls  \emph{only the strength of the effective interaction} while the spatial extension is unaffected. We find that the strongest effective interaction is reached when the coupling constants $g_j^{(ab)}$ have opposite sign.

Since we are interested only in  the effective interaction  between the impurities mediated by the BEC mixture, in our analysis (Fig. \ref{fig:Energia2ImpurezzeMixRb87}) we have not included the interaction energy due to the direct  overlap between the impurity wavefunctions when they are in the same lattice site .


\subsection*{Polarons in Presence of Raman Coupling}
A much richer physics appears when one introduces a Raman coupling between the two hyperfine levels $\ket{a}$ and $\ket{b}$, Fig. \ref{fig:RamanTransition},
described by the extra Hamiltonian Raman term of equation \eqref{eq:LaserRamanHam}. 
Indeed is clear from equation \eqref{eq:CoeffabcLaser} that this affects the BEC coupling constants $g_{ij}^{(b)}$, introducing a further parameter for tuning the effective interaction between polarons. 

\begin{figure}[h]
\centering
\includegraphics[width= 0.33\columnwidth]{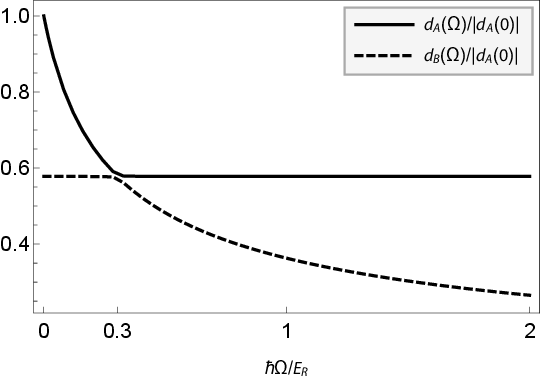} 
\includegraphics[width= 0.33\columnwidth]{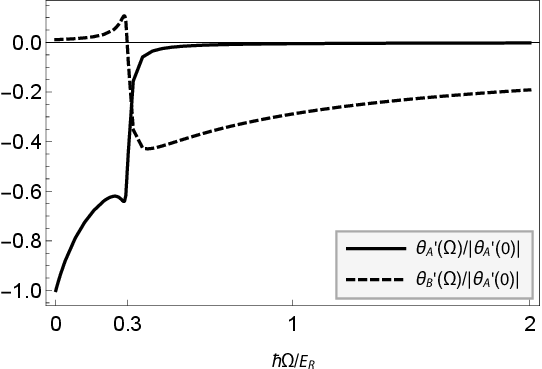} 
\includegraphics[width= 0.33\columnwidth]{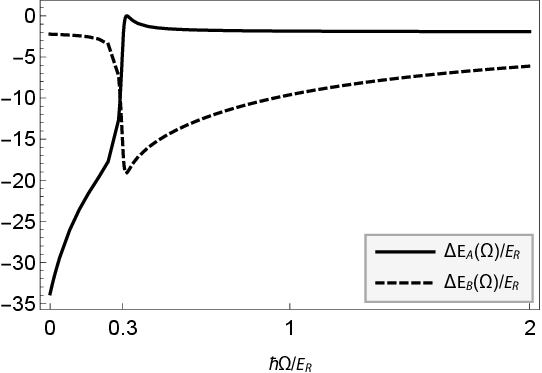} 
\caption{\textbf{Raman controlled effective deformations of the BEC order parameter}: 
width of the effective healing lengths $d_j(\Omega)=1/\eta_j(\Omega)$, normalised for the value in absence of Raman coupling (left). 
Depth of the effective deformations as a function of the Raman coupling (centre).
Contribution of each effective deformation to the system energy$\Delta E_i(\Omega)$, equation \eqref{eq:TwoImpuritiesRelEnergy}, in recoil energy units $E_R$ (right). 
The value of the threshold Raman coupling, equation \eqref{eq:OmegalimiteRaman} is $\hbar\Omega_{\text{lim}}\simeq 0.61 E_R$. 
(We have assumed the following values: $g_A^{(ab)}=-2 g_B^{(ab)}$ where $g_B^{(ab)}=10 g_{AA}^{(b)}$ and relative distance between impurities set to $d=0$.) 
}
\label{fig:EffDeformationRaman}
\end{figure}

A first clear effect of the Raman coupling is to modify both the depths and the widths of the effective deformations. The widths change as $d_i {=} 1/\eta_i$, where $\eta_i$ are the effective Raman coupling dependent healing lengths in \eqref{eq:MEtaAEtaBnAnB}
\begin{equation}
\eta_\pm(\Omega)= \left\{\left(\frac{2 m_b}{\hbar^2}\right) \left[\left( g_{AA}^{(b)}+g_{BB}^{(b)}\right)n \ + \frac{\hbar\Omega}{2}\pm \sqrt{\left(g_{AA}^{(b)}-g_{BB}^{(b)}\right)^2 n^2+\left(\frac{\hbar\Omega}{2}-g_{AB}^{(b)} n\right)^2}\ \right]\right\}^{1/2}
 \label{eq:etaaetabLasernmezzi}
\end{equation}

which show a clear dependence on the Rabi coupling strength $\Omega$. For $\Omega{\rightarrow} 0$ we obtain the same result as in (\ref{eq:MEtaAEtaBnAnB}), while for strong coupling $\Omega{\gg} \vert g_{ij} n\vert$ we find
\begin{align}
d_+&{\sim_{+\infty}} \left[
\frac{\hbar^2 / 2m_b}{\left(g_{AA}^{(b)}+g_{BB}^{(b)}+g_{AB}^{(b)}\right)n}\right]^{1/2},
& 
d_-&{\sim_{+\infty}}
\sqrt{ \frac{\hbar}{2m_b \Omega}}
\end{align}
This means that above a threshold value $\Omega_{lim}$ of the laser Rabi frequency the width of one effective deformation tends to a constant value while the width of other one goes to zero. The threshold value does not depend on the impurity atoms parameters and is equal to
\begin{equation}
\frac{\hbar\Omega_{lim}}{2}=\left[\frac{\left(g_{AA}^{(b)}-g_{BB}^{(b)}\right)^2 +\left(g_{AB}^{(b)}\right)^2}{2 g_{AB}^{(b)}}\right]n
\label{eq:OmegalimiteRaman}
\end{equation}
In our system we obtain $\Omega_{lim}/2\pi{\simeq} 615$Hz. In Fig. \ref{fig:EffDeformationRaman} we show how the effective deformations' widths $d_i{=}1/\eta_i(\Omega)$  as a function of the Raman coupling strength ( equation \eqref{eq:etaaetabLasernmezzi}). 
To find the effect of the Raman coupling on the BEC deformations we proceed, as before, 
by solving the system (\ref{eq:MatrixFormSysDef2BECModel}) with  coefficients (\ref{eq:CoeffabcLaser}) but now with $\Omega{\neq}0$. 

For two BECs with the same density $n_{0i}{=}n/2$, in the 1D case, the effective deformations become $\theta_i'(x){=} \mathcal{K}_{i}^{las} \mathcal{F}_{\sigma,\eta_i(\Omega)}(x)$
where the $\mathcal{F}_{\sigma,\eta_i}(x)$ and the constants $\mathcal{K}_{i}^{las}$ are given explicitly in equation \eqref{eq:FunzioniFDeforModelAppA} and equation \eqref{eq:LasKConstants}, which depend both on the impurity atoms and the BEC coupling constants. 
We show in Fig. \ref{fig:EffDeformationRaman} how the effective deformations depths change due to the Raman coupling. 
The ground state energy of the system is obtained with the same technique discussed above.
For a single impurity immersed in a deep 1D optical lattice we find
\begin{equation}
E_{GP}^{Ram}(\Omega)=E^{one}_{GP}(\Omega)-\left(\frac{\hbar\Omega}{2}\right)\frac{1}{2\eta_{-} \eta_{+}\left(\eta_{-}^2-\eta_{+}^2\right)}\left[e^{\frac{1}{2} \sigma ^2 \eta_{+}^2} \text{Erfc}\left(\frac{\sigma  \eta_{+}}{\sqrt{2}}\right) \eta_{-}-e^{\frac{1}{2} \sigma^2 \eta_{-}^2} \text{Erfc}\left(\frac{\sigma  \eta_{-}}{\sqrt{2}}\right) \eta_{+}\right]
\label{eq:EnSingolaLaser}
\end{equation}
where $\eta_i{\equiv}\eta_i(\Omega)$ (see equation \eqref{eq:etaaetabLasernmezzi}). 
For a system of two impurities the ground state energy as function of their relative distance is (we set the zero of energy for $d{\rightarrow}\infty$):
\begin{equation}
\Delta E^{Ram}(d,\Omega)=\sum\limits_{i=\pm}\Delta E_i(d,\Omega)-\frac{\hbar\Omega}{2}\mathcal{Q}_{\sigma,\eta_{-}(\Omega),\eta_{+}(\Omega)}(d)
\label{eq:TwoImpuritiesRelEnergy}
\end{equation}
where $\Delta E_i(d,\Omega)$ are obtained from the equation \eqref{eq:EnergyTwoImpDistance} with the substitution $\eta_i{\rightarrow}\eta_i(\Omega)$ defined in equation \eqref{eq:etaaetabLasernmezzi} while the term $\mathcal{Q}_{\sigma,\eta_A(\Omega),\eta_B(\Omega)}(d)$ represents the direct coupling of the condensate deformations due to the laser. This latter term, whose explicit form is shown in equation \eqref{eq:IntegrModelQFunction} of Methods and equation \eqref{eq:IntegrModelQFunction2}, turns out to be negligible compared to $\Delta E_i (d,\Omega)$. In Fig. \ref{fig:EffDeformationRaman} (right) the two energy contribution $\Delta E_i$ in equation  \eqref{eq:TwoImpuritiesRelEnergy} are shown as function of the Raman coupling, in energy recoil units, for relative distance $d=0$ and $g_B^{(ab)}=10 g_{AA}^{(b)}$,  $g_A^{(ab)}=-2 g_B^{(ab)}$. It can be seen clearly that the effect of the Raman term is to modulate the effective interaction between the two impurity atom, which decreases for large values of $\Omega$. 

\begin{figure}[h]
\centering
\includegraphics[width= 0.3\columnwidth]{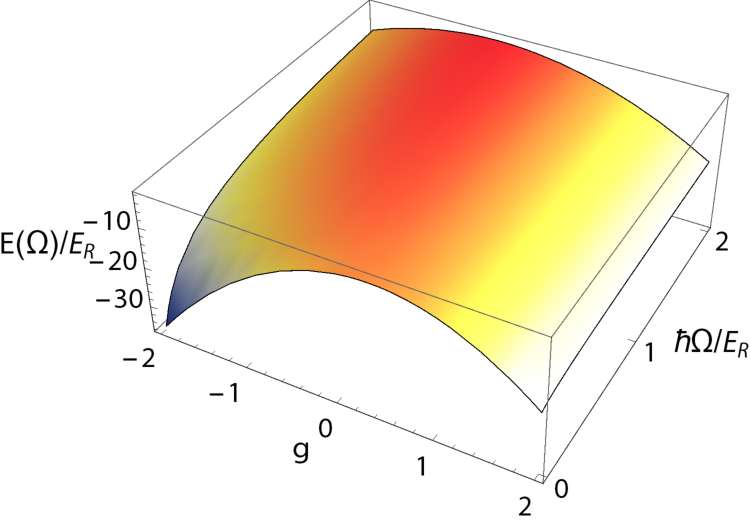} 
\includegraphics[width= 0.3\columnwidth]{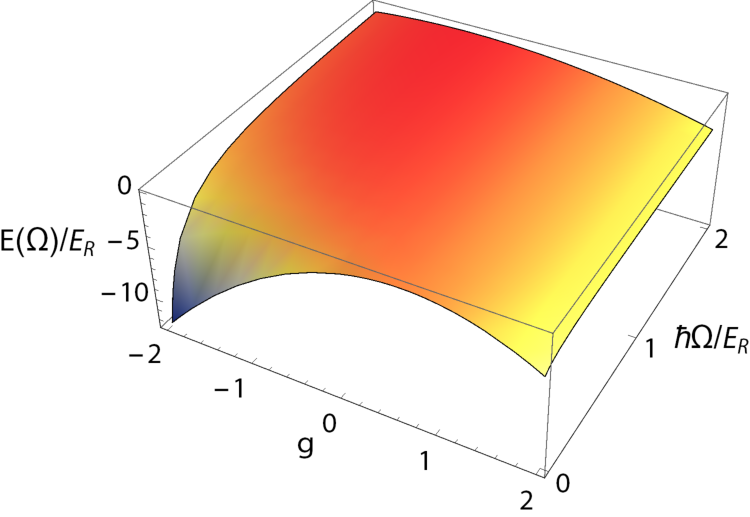} 
\includegraphics[width= 0.3\columnwidth]{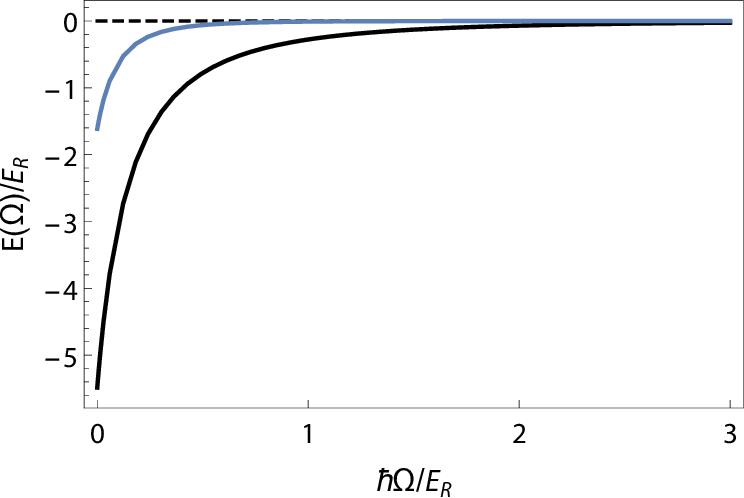} 
\caption{\textbf{Raman controlled effective interaction between two impurity atoms }: $\Delta E^{Ram}(d,\Omega)$ as function of the Raman laser coupling $\hbar\Omega$ and of the ratio of the impurity-BEC coupling constants $g = g_{A}^{(ab)}/g_{B}^{(ab)}$ and $g_{B}^{(ab)} = 10 g_{AA}^{(ab)}$. The Raman strength is in recoil energy units. The relative distance $d{=}\vert x_1 {-} x_2\vert$ between the impurities is fixed to  $d=0$ (left), $d=a$ (centre), where $a=\lambda/2$ is the lattice spacing. (right) $\Delta E^{Ram}(d,\Omega)$, respectively for $d=a$ (black line) and $d=2a$  (blue line)with $g=-1$, as a function of the Raman coupling strength.}
\label{fig:EnTwoImpuRaman}
\end{figure}

In Fig. \ref{fig:EnTwoImpuRaman} we study in detail how the Raman strength influences the effective interaction between two impurity atoms, for fixed value of the ratio between the coupling constants $g=g_{A}^{(ab)}/g_{B}^{(ab)}$. We show in Fig. \ref{fig:EnTwoImpuRaman} (left) that if the distance between the impurities is fixed to $d=0$, the Raman term allows one to modulate the effective interaction in a wide range. More remarkably, when the two impurities are placed in different sites, the Raman coupling allows one to completely turn off the effective polaron mutual interaction. As shown in Fig. \ref{fig:EnTwoImpuRaman} (centre) the effect is more pronunced when the impurity-BEC coupling constants $g_j^{(ab)}$ have opposite sign. Specifically in Fig. \ref{fig:EnTwoImpuRaman} (right) we plot the effective interaction for two impurities with relative distance of one lattice site $d=a$ (black line), and two sites $d=2a$ (blue line), for $g=g_{A}^{(ab)}/g_{B}^{(ab)}=-1$.

Therefore by a suitable choice of the Raman coupling i.e. of the external laser field intensity, one can tune the polaron effective interaction potential from attractive to zero. Such behavior is due to the combined effect of the increase of the density and the reduction of the size of the effective deformations near the position of the impurity atoms. 

\begin{figure}[h]
\centering
\includegraphics[width= 0.45\columnwidth]{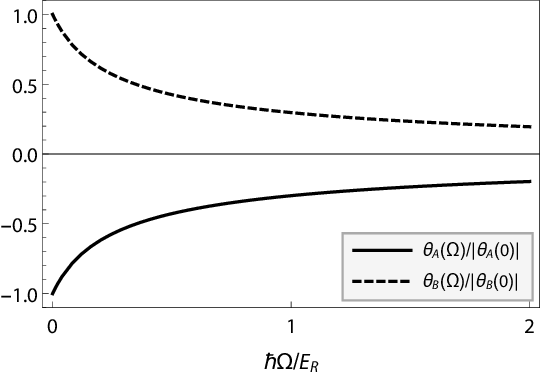}
\includegraphics[width= 0.45\columnwidth]{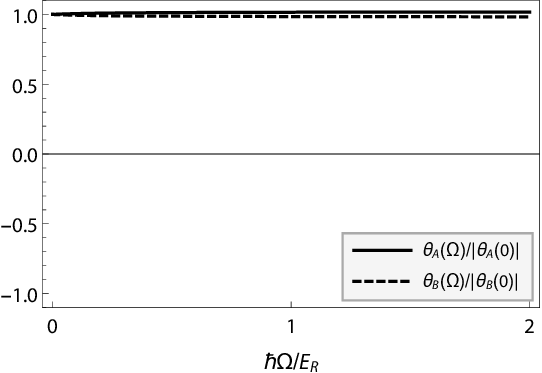} 
\caption{\textbf{Effect of the Raman coupling on the real deformations}: Depth of the real deformations $\theta_j(x=0,\Omega)$, in the position of the impurity, $x=0$, as a function of the Raman coupling $\Omega$ for (left)
$g_A=-g_B=-10 g_{AA}^{(ab)}$  and for (right) $g_A=g_B=10 g_{AA}^{(ab)}$ . 
}
\label{fig:PlotRamanRealDeformations}
\end{figure}

The laser term changes the effect of the coupling constants $g_{ij}^{(b)}$ on the effective deformations of the BEC as it can be seen in equation  \eqref{eq:CoeffabcLaser}, which consequently allows tuning the interaction between impurities. 
To better understand the physics underlying our scheme, we evaluate the effect of the Raman coupling on the real deformations $\theta_A(x),\theta_B(x)$ by using equation \eqref{eq:EffDeformTransform}. In Fig. \ref{fig:PlotRamanRealDeformations} we plot the depth of the BEC deformations $\theta_j(x=0,\Omega)$ as a function of the Raman strength $\Omega$ in the position of the impurity $x=0$. We observe that if the coupling constants between impurity and the two BEC component have opposite sign (Fig. \ref{fig:PlotRamanRealDeformations} (left)) the Raman term reduces the depth of the deformations, which effectively allows a tunability of the polaron binding energy and of the effective interaction between two impurities immersed in the system. On the other hand, if the coupling constants have the same sign, Fig. \ref{fig:PlotRamanRealDeformations} (right), the Raman term has a weaker effect on the tunability of the system. 

\section*{Conclusion}
We have analysed the properties of polarons which originate from the coupling of   atomic impurities with a two component BEC. 
We show how the polaron-polaron interaction can be tuned by acting on the coupling constant values between impurities and a two component BEC. 

Moreover we find that a suitable Raman coupling between the hyperfine levels of the BEC allows tuning the effective potential which describes the polaron-polaron interaction.  Specifically we found that the range of tunability depends on the ratio between the coupling constants between impurity and BEC, which is maximised when the coupling with the two BEC component has opposite sign. More remarkably we find that the Raman term can switch the effective interaction from attractive to zero for impurities in different sites. 
Our proposal is within the state of the art of experimental realisation and could be probed by  using radio-frequency spettroscopy techniques \cite{hu_bose_2016-1,jorgensen_observation_2016-1,schirotzek_observation_2009,zhang_polaron--polaron_2012}, already employed for impurity atoms in a single BEC environment. 

Our study opens the way to new exciting questions in the controlled many-body dynamics and thermodynamics of ultracold atomic mixtures, including issues of thermalisation, dissipative preparation of strongly-correlated states and observation of new exotic phases in systems of many polarons.

\section*{Methods}
The system Hamiltonian is $\hat H {=} \hat H_a {+} \hat H_b {+} \hat H_{ab}$ where $a$ and $b$ label respectively the impurity and the BEC atoms and where
\begin{align}
\hat H_a &=\int d \g{r}\ \hat \chi^\dag(\g{r}) \left( -\frac{\hbar^2 \nabla^2}{2 m_a}+V(\g{r})-\mu^{(a)}+ g^{(a)} \hat \chi^\dag (\g{r})\hat \chi(\g{r})\right)\hat \chi(\g{r})\nonumber \\
\hat H_b &=\sum_{i=A,B}\int d\g{r}\ \hat \psi_i^\dag (\g{r})\left[-\frac{\hbar^2 \nabla^2}{2 m_b}+U(\g{r})-\mu_i^{(b)} +\right. \nonumber \left.  g_{ii}^{(b)} \hat \psi_i^\dag (\g{r})\hat \psi_i (\g{r}) +\sum_{i<j} g_{ij}^{(b)} \hat \psi_j^\dag(\g{r})\hat \psi_j(\g{r})\right] \hat \psi_i (\g{r})
  \nonumber\\
\hat H_{ab}&=\sum_{i=A,B} \int d \g{r}\ g_i^{(ab)} \hat \chi^\dag(\g{r}) \hat\psi_i^\dag (\g{r}) \hat \chi(\g{r}) \hat\psi_i (\g{r})
\label{eq:HamBECsModel1}
\end{align}
Here $\hat{\chi}(\g{r})$ and $\psi_i^\dag (\g{r})$  are respectively the impurity field operator and the BEC field operators, with $i{=}A,B$ labelling the two BEC component and $ g_{ij}^{(b)}$  and $g_{i}^{(ab)}$ are the interspecies and intraspecies pseudo-potential coupling constants. The chemical potentials of the impurity and the BECs are labeled respectively with $\mu^{(a)}$ and $\mu_i^{(b)}$. 

In an optical potential, once the impurity field operator is expressed in terms of Wannier functions $\chi_i(\mathbf{r})$ centred around the lattice site $i$ as  $\hat \chi(\g{r}){=} \sum_i \chi_i(\g{r}) \hat a_i $, the impurities' Hamiltonian $\hat H_a$ is described by a Bose-Hubbard model \cite{lewenstein_ultracold_2012}.  For a deep lattice potential impurities are described by the average density $\rho(\g{r}){=}\langle \hat \chi^\dag(\g{r})\hat \chi^\dag(\g{r})\rangle{=}\sum_i n_i \vert \chi_i(\g{r})\vert^2$, 
where  $n_i$ is the average number of impurities in site $i$, and the Wannier functions are gaussians \cite{bruderer_polaron_2007,bruderer_transport_2008}.

A static polaron is obtained with a deep optical potential in which the hopping dynamics is suppressed \cite{bruderer_polaron_2007,bruderer_transport_2008}. A mean field description of the two component BEC is formally obtained substituting the field operator $\hat \psi_i(\mathbf{r})$ with the unperturbed order parameters plus a linear correction due to interaction with impurities $\hat \psi_{i}(\mathbf{r}){\rightarrow}  \psi_{0i}(\mathbf{r}){+}\theta_i(\mathbf{r})$ \cite{bruderer_polaron_2007,bruderer_transport_2008}. For real $\psi_{0i}(\g{r})$ ed $\theta_i(\g{r})$ we find 
\begin{align}
H_{GP}&=\sum_{i=A,B}\int d\g{r}\ \left\{
\psii(H_0 - \mu_i^{(b)})\psii + g_{ii}^{(b)} \psi_{0i}^{4}(\g{r}) + \right. \nonumber \left.  \sum_{i<j} g_{ij}^{(b)} \psi_{0i}^2(\g{r})\psi_{0j}^2(\g{r})+\sum_{i=A,B}g_i^{(ab)} \rho(\g{r}) \psi_{0i}^2(\g{r})\right\}+\nonumber  \\ 
&+\sum_{i=A,B}\int d\g{r}\ g_{i}^{(ab)} \rho(\g{r})\psii\thi 
\label{eq:HGPStazCondModel}
\end{align}
where $H_0{\equiv}-(\hbar^2\nabla^2)/(2m_b){+}U(\mathbf{r})$. For a shallow trap $U(\mathbf{r}){=}0$ and the unperturbed BEC wave functions $\psi_{0i}(\mathbf{r})$ are constant and real \cite{fetter_nonuniform_1972}. Using the stationarity condition $\delta H_{GP}/\delta\theta_{0i}{=}0$ we obtain a  system of coupled differential equationsfor the deformations $\theta_i (\mathbf{r})$  (Eq.~(3) in the main text ). Defining $\mathbb{S}$ the transformation matrix that diagonalises $\mathbb{M}$ in \eqref{eq:CoeffabcLaser} we obtain a decoupled system of two modified Helmholtz equations in the base of effective deformations
\begin{equation} 
(\theta'_A,\theta'_B)^T{=}\mathbb{S}^{-1}\left(\theta_A,\theta_B\right)^T
\label{eq:EffDeformTransform}
\end{equation}
The explicit expressions for the effective deformations in a one-dimensional two component BEC with no Raman coupling are $\theta_i'(x){=}\mathcal{K}_{\pm} \mathcal{F}_{\sigma,\eta_i}(x)$
where
\begin{equation}
\begin{split}
\mathcal{F}_{\sigma,\eta_i}\equiv
\frac{1}{4\eta_i} &\left\{\exp\left[\frac{1}{4} \eta_i  \left(\eta_i  \sigma ^2+4 x\right)\right] \text{Erfc}\left(\frac{\eta_i  \sigma }{2}+\frac{x}{\sigma }\right) +\right. \left. \exp\left[\frac{1}{4} \eta_i  \left(\eta_i  \sigma ^2-4 x\right)\right] \text{Erfc}\left(\frac{\eta_i  \sigma }{2}-\frac{x}{\sigma }\right)\right\}
\end{split}
\label{eq:FunzioniFDeforModelAppA}
\end{equation}
and
\begin{equation}
\mathcal{K}_{\pm}=\pm \left(\frac{m_b}{\hbar^2}\right)\sqrt{\frac{n}{2}} \left[\frac{g_{A}^{(ab)} g_{AB}^{(b)} + g_{B}^{(ab)}\left( g_{AA}^{(b)}-g_{BB}^{(b)} \mp \beta_0\right)}{\beta_0}\right]
\label{eq:KConstantsNoLaser}
\end{equation}
with $\beta_0{\equiv} \left[(g_{AA}^{(b)}-g_{BB}^{(b)})^2 + (g_{AB}^{(b)})^2\right]^{1/2}$. The ground state energy of a system of impurities interacting with the two component BEC is obtained using \eqref{eq:EffDeformTransform} in \eqref{eq:HGPStazCondModel}. In particular, taking into account only terms containing $\rho(\mathbf{r})$, for BECs with the same density, $n_{0i}{=}n{/}2$ we obtain the following value for the coefficients $A_0$ and $B_i$ 
\begin{align}
A_0&=\frac{n}{2}\left(g_A^{(ab)}{+}g_B^{(ab)}\right), &
B_{i}&=\sqrt{\frac{n}{2}} \left[k_i g_A^{(ab)}{+} g_B^{(ab)}\right]L_i
\label{eq:EGPConstantiMoltipModel}
\end{align}
Here we define $k_A{\equiv} k_-$ and $k_B{\equiv} k_+$, where we introduce the constants
$k_\pm{\equiv}\left(g_{AA}^{(b)}-g_{BB}^{(b)}\pm \beta_0\right)/g_{AB}^{(b)}$ and \\ $L_i{\equiv} \sqrt{n/2}\left(m_b g_{AB}^{(b)}/\hbar^2 \beta_0\right)\left(g_{A}^{(ab)}-k_{i\neq j} g_{B}^{(ab)}\right)$.

\subsection*{Mean-Field Description With Raman Term}
\begin{figure}[t]
\centering
\includegraphics[width=0.4\textwidth]{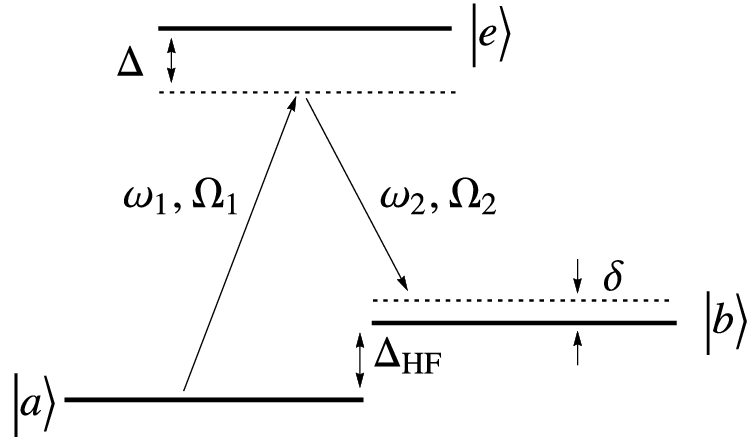} 
\caption{\textbf{Raman coupling in a $\Lambda$ system}: three hyperfine levels of $^{87}Rb$ are coupled by two laser fields with frequency $\omega_1$, $\omega_2$ and Rabi frequency $\Omega_1$ and $\Omega_2$. Here $\Delta$ represents the common detuning with the state $\ket{e}$, while $\Delta_{HF}$ is the energy difference between the hyperfine levels $\ket{a}$ e $\ket{b}$.}
\label{fig:RamanTransition}
\end{figure}

We introduce an external coupling term between the two BECs hyperfine level in $\ket{a}$ and $\ket{b}$ the form of Raman coupling as shown in Fig. \ref{fig:RamanTransition}. 
Here $\omega_i$ and $\Omega_i$ are respectively the laser frequency and the Rabi frequency. For a large enough detuning $\Delta$ the hamiltonian can be cast in the following mean field effective hamiltonian by adiabatically eliminating the excited level $\ket{e}$ \cite{kanamoto_quantum_2007
,brion_adiabatic_2007,cirac_quantum_1998}. 
When $\omega_2{=}\omega_1$ the two BECs' population remain constant in time \cite{cirac_quantum_1998,jaksch_cold_2005}, and defining  the Raman coupling strength $\Omega{\equiv} 4\Omega_1\Omega_2/\Delta$ we have
\begin{align}
&\hat H_{Ram}=-\frac{\hbar\Omega}{2}\int d\g{r}\ \left( \hat \psi_A^\dag(\g{r}) \hat \psi_B(\g{r}) + \hat \psi_B^\dag (\g{r})\hat \psi_A(\g{r})\right)
\label{eq:LaserRamanHam}
\end{align}
Hence the total hamiltonian of the system  is $\hat H{=}\hat H_a {+} \hat H_b {+} \hat H_{ab}{+}\hat H_{Ram}$.
Separating the contributions describing the uniform unperturbed condensate and the deformation induced by the impurity we find the mean field description adding the term
\begin{equation}
\begin{split}
H_{GP}^{Ram}&=\frac{\hbar\Omega}{2}\sum_{i<j}\int d\g{r}\ \left[\psi_{0i}^*(\g{r})\psi_{0j}(\g{r})+\psi_{0i}(\g{r})\psi^*_{0j}(\g{r})\right]-\frac{\hbar\Omega}{2}\sum_{i<j}\int d\g{r}\ \left[\theta_i^*(\g{r})\theta_j(\g{r})+\theta_j^*(\g{r})\theta_i(\g{r})\right]
\end{split}
\label{eq:HamLaserHGPModel}
\end{equation}
to the mean field description \eqref{eq:HGPStazCondModel}. 
The effective deformations with the Raman coupling term are $\theta_i'(x){=}\mathcal{K}_{\pm} ^{las}\mathcal{F}_{\sigma,\eta_i}(x)$, where
\begin{equation}
\begin{split}
&\mathcal{K}^{las}_{\pm}=\pm \left(\frac{m_b}{\hbar^2}\right)\sqrt{\frac{n}{2}}
 \left[\frac{g_{A}^{(ab)} \left(\hbar \Omega-2  g_{AB}^{(b)} n \right) + g_{B}^{(ab)}\left( 2 n \left(g_{AA}^{(b)}-g_{BB}^{(b)}\right)\mp \beta\right)}{\beta}\right]
\end{split}
\label{eq:LasKConstants}
\end{equation}
and $\beta{\equiv}\sqrt{4 \left(g_{AA}^{(b)}-g_{BB}^{(b)}\right)^2 n^2
+\left( \hbar\Omega-2 g_{AB}^{(b)} n\right)^2}$. As shown in the main text we find the ground state energy as function of the relative distance and the Raman coupling strength $\Omega$ for a system of two impurities in a deep 1D optical lattice. Explicitly, assuming gaussian Wannier wavefunctions, we solve the integral
\begin{equation}
\begin{split}
&\mathcal{D}_{\sigma,\eta_A,\eta_B}(d)\equiv \frac{1}{4\eta_A \eta_B \pi \sigma^2}\int dxdx' dx''\ e^{-\eta_A \vert x-x'\vert}e^{-\eta_B \vert x-x''\vert}e^{-(x')^2/\sigma^2}e^{-(x''-d)/\sigma^2}
\end{split}
\label{eq:IntDueImpurezze}
\end{equation}
The result for $\eta_B<\eta_A$ is:
\begin{equation}
\begin{split}
&\mathcal{D}_{\sigma,\eta_A,\eta_B}(d)=
\frac{1}{4\eta_A \eta_B (\eta_B-\eta_A)(\eta_B+\eta_A)}
\left\{\eta_B e^{-\frac{1}{2} \eta_A \left(2 d+\eta_A \sigma ^2\right)} \left[\left(e^{2 d \eta_A}+1\right)+\text{Erf}\left(\frac{d-\eta_A \sigma ^2}{\sqrt{2} \sigma }\right)-e^{2 d \eta_A} \text{Erf}\left(\frac{d+\eta_A \sigma ^2}{\sqrt{2} \sigma }\right)\right]\right.\\
&\left. -
\eta_A e^{\frac{1}{2} \eta_B \left(\eta_B \sigma ^2-2 d\right)} \left[\left(e^{2 d \eta_B}-1\right)+\text{Erf}\left(\frac{d-\eta_B \sigma ^2}{\sqrt{2} \sigma }\right)-e^{2 d \eta_B} \text{Erf}\left(\frac{d+\eta_B \sigma ^2}{\sqrt{2} \sigma }\right)\right]\right\}
\end{split}
\label{eq:IntegrModelQFunction}
\end{equation}
For $\eta_A =\eta_B=\eta$ the result of the integral \eqref{eq:IntDueImpurezze} is:
\begin{equation}
\begin{split}
\mathcal{D}_{\sigma,\eta,\eta}(d)&=\frac{\sigma}{2\sqrt{2\pi}\eta^2}e^{-d^2/2\sigma^2}+
\frac{1}{8\eta^3}\left\{e^{\frac{\eta}{2}(-2 d +\eta\sigma^2)}\left(\eta(d-\eta\sigma^2)+1\right)\left[1+\text{Erf}\left(\frac{d-\eta\sigma^2}{\sqrt{2}\sigma}\right)\right]-\right.\\
&\left.e^{2 d \eta}\left(\eta(d+\eta \sigma^2)-1\right)\left[1-\text{Erf}\left(\frac{d+\eta\sigma^2}{\sqrt{2}\sigma}\right)\right]\right\}
\label{eq:IntegrModelQFunction2}
\end{split}
\end{equation}

\noindent 


\section*{Acknowledgements}
This work is supported by the John Templeton Foundation (grant ID 43467) and the EU Collaborative Project TherMiQ (Grant Agreement 618074). GMP thanks the MIUR for support under PRIN 2010/11. EC is currently supported by European Research Council under the European Union's Seventh Framework Programme (FP/2007-2013) / ERC Grant Agreement No. 308253. The authors thank Pietro Massignan, Gerald Milburn, Tomi H. Johnson and Luca Marmugi for interesting discussions and suggestions.

\section*{Author contributions statement}
E.C. performed all the calculations under the technical help of G.D.C. and with comments and inputs by all the authors. E.C., G.D.C. and G.M.P wrote the manuscript with feedback from D.A. . All the authors contribute to the analysis of the results and to the review of the manuscript. G.M.P. planned and supervised the project. 
\section*{Additional information}
\textbf{Competing financial interest: } The authors declare no competing financial interests. 

\section*{Supplementary Materials}
\subsection*{Excitations of the system}
In the main text the system has been analysed for zero temperature, namely in absence of excitations of the BEC mixture. However the static polaron description can be further extended even for $T{\neq}0$. In this section we show that the excitation spectrum is independent from the presence of impurity atoms even for a mixture of BEC. This means that the description of our system in presence of excitations (i.e. when excitation modes are populated by temperature effects) is of static polarons and non-interacting quasi-particle excitations. This result has been proved for impurity atoms in one and two bands of a lattice, interacting with a single BEC in \cite{Bruderer2008a,Yao2015}. 

To find the excitation spectrum we use a standard approach, namely taking into account small excitations $\hat \zeta_i(x)$ around the ground state of the system. Formally we operate the substitution on the BEC field operators $\hat \Psi_i(x){\rightarrow}\psi_{0i}(x)+\theta_i(x)+\hat \zeta_i(x)$ up to second order terms in $\hat \zeta_i(x)$ in the Hamiltonian $\hat H{=}\hat H_a {+} \hat H_b {+} \hat H_{ab}{+}\hat H_{Ram}$, 
Eq.~(12) and Eq.~(16). First order terms are zero due to stationary conditions.  Firstly we consider the case of no Raman laser coupling in Eq.~(12).
The excitations' Hamiltonian of a system of two BEC interacting with impurity atoms is $\hat H_\zeta$ where 
\begin{equation}
\begin{split}
&\hat H_{\zeta}=
\sum_{i=A,B}\int dx\ \zic\left[H_0-\mu_{i}^{(b)}+ 
g_{ii}^{(b)}4\vert\psii\vert^2 +\right. \left. \sum_{i<j} g_{ij}\vert\psi_{0j}^2(x)\vert^2\right]\zi+\\
&+\sum_{i=A,B}\int dx\ g_{ii}^{(b)}\left[\psi_{0i}^{*2}(x)\zi\zi+H.C.\right]+\\
&+\sum_{i<j}\int dx\ g_{ij}^{(b)} 
\left[\psii\psijc\zic\zj+\right. \left. \psii\psij\zic\zjc + H.C.\right]
\end{split}
\label{eq:Hzeta}
\tag{S1}
\end{equation} 
We underline in particular that the latter is independent from impurity atoms and therefore the excitation spectrum is independent from impurities'  presence. The Hamiltonian \eqref{eq:Hzeta} is a quadratic form in the operator $\hat \zeta_i(x)$ and therefore it can be reduced to a diagonal form $\hat H_\zeta{=}\sum_n\sum_{k\neq 0} \hbar\omega_{n k} \hat b_{n k}^\dag \hat b_{n k}$ by Bogoliubov transformations, after rewriting it in terms of plane waves:
\begin{equation}
\zi=\frac{1}{\sqrt{V}}\sum_{\g{k}} e^{i\g{k}\cdot\g{r}}\ \hat c_{ik}  
\tag{S2}
\end{equation}
where we define the quasi-particle operators:
\begin{equation}
\hat b_{n k} = \sum_{i} \left[u_{n i}(k)\hat c_{i k}+v_{n i}(k)\hat c_{i-k}^\dag\right]\qquad\qquad i \in \left\{A,B\right\}
\label{eq:BogoliubovOperatorTransfModel}
\tag{S3}
\end{equation}
as shown in \cite{searchrojo2001,GolsteinMoore2000}. The effect of the operators $\hat b_{n k}$ is to annihilate a quasiparticle which is a collective excitation of the system. The excitation spectrum has two branches $\hbar\omega_\pm$ and explicitly, for BEC with the same density $n_{0i}{=}n{/}2$, it is
\begin{equation}
\begin{split}
&\hbar\omega_\pm(k)=
\left\{
\left(\frac{\hbar^2 k^2}{2 m_b}\right)
\right. \left.
 \left[\left(\frac{\hbar^2 k^2}{2 m_b}+(g_{AA}^{(b)}+g_{BB}^{(b)})n\right) \pm n\sqrt{\left(g_{AA}^{(b)} -g_{BB}^{(b)}\right)^2 + \left(g_{AB}^{(b)}\right)^2 }\right]
\right\}^{1/2}
\end{split}
\label{eq:SpectrumBogolon2BEC}
\tag{S4}
\end{equation}
For little $k$ the excitation energy depends linearly on the momentum as $\hbar\omega(k){=}v_s \hbar k  {+} O(k^3)$, that means the excitations are phonon-like and $v_s$ is the sound velocity of the BEC mixture 
\begin{equation}
v_s^\pm=\left\{
\frac{n}{2 m_b}\left[g_{AA}^{(b)}+g_{BB}^{(b)}\pm \sqrt{\left(g_{AA}^{(b)} -g_{BB}^{(b)}\right)^2 + \left(g_{AB}^{(b)}\right)^2 }\right]
\right\}^{1/2}
\tag{S5}
\end{equation}
When the Raman laser coupling is included in the system the excitation Hamiltonian \eqref{eq:Hzeta} becomes
\begin{equation}
\hat H_\zeta^{Ram}=\hat H_\zeta - \frac{\hbar\Omega}{2}\sum_{i<j}\int dx\ \left[\hat \zeta_i^\dag(x) \hat \zeta_j(x)+ \hat \zeta_j^\dag (x)\hat \zeta_i(x)\right]
\tag{S6}
\end{equation} 
Following \cite{Book:Meystre} we evaluate the excitation spectrum when the Raman coupling term is included in the system. The explicit expression is quite complex and here we show it in the limit $g_{AA}^{(b)}{=}g_{BB}^{(b)}$ and $n_{0i}{=}n/2$. Explicitly
\begin{align}
\hbar\omega_{-}(k)&=\sqrt{\left(\frac{\hbar^2 k^2}{2 m_b}+ \hbar \Omega\right) \left[\frac{\hbar^2 k^2}{2 m_b}+\hbar \Omega+  \left(2 g_{AA}^{(b)} -g_{AB}^{(b)}\right)n \right]}\nonumber \\
\hbar\omega_{+}(k)&=\sqrt{\frac{\hbar^2 k^2}{2 m_b} \left[\frac{\hbar^2 k^2}{2 m_b}+\left(2 g_{AA}^{(b)}+g_{AB}^{(b)}\right)n \right]}
\tag{S6}
\label{eq:2Branches1}
\end{align}
Finally we find that the excitation spectrum in our system is unaffected by the presence of impurity atoms. We highlight that the description of our system in presence of excitations (i.e. when excitation modes are populated by temperature effects) is then of static polarons and quasi-particles with the energy spectrum in \eqref{eq:2Branches1}. 

It is interesting to analyse the excitation spectrum in the limit case of $k\rightarrow 0$, where we find that the expressions of the low excitation spectrum are
\begin{align}
\hbar \omega_{-} (k)&= \sqrt{\hbar \Omega \left(\hbar \Omega+(2 g_{AA}^{(b)}-g_{AB}^{(b)})n\right)}+\frac{\hbar^2 k^2}{2 m_b}\left[\frac{(2 g_{AA}^{(b)}-g_{AB}^{(b)})n +2 \hbar\Omega }{2\sqrt{\hbar\Omega\left(\hbar\Omega+(2 g_{AA}^{(b)}-g_{AB}^{(b)})n\right)}}\right] +\mathcal{O}(k^2)\nonumber\\
 \hbar \omega_{+} (k)&= \sqrt{\frac{(2 g_{AA}^{(b)}+g_{AB}^{(b)})n}{2 m_b}}\hbar k +\mathcal{O}(k^2)
\tag{S7}
\end{align}
Regarding the low excitation sector from the latter we see that when the two BEC are Raman coupled the branch $\hbar \omega_{-}(k)$ is no longer phononic and it is characterised by an energy gap \cite{Book:Meystre}, while the other is unaffected by the laser coupling term and depends only on the physical characteristics of the BEC. On the other hand, when the laser term $\hbar\Omega$ is off we see from Eq. \eqref{eq:2Branches1} that both branches are phononic with a different sound speed: 
\begin{align}
\hbar \omega_{-} (k)&= \sqrt{\frac{(2 g_{AA}^{(b)}-g_{AB}^{(b)})  n}{2 m_b}}\hbar k +\mathcal{O}(k^2)
\nonumber\\
\hbar \omega_{-} (k)&= \sqrt{\frac{(2 g_{AA}^{(b)}+g_{AB}^{(b)}) n}{2 m_b}}\hbar k +\mathcal{O}(k^2)
\tag{S8}
\end{align}

\end{document}